\documentclass[11pt,a4paper]{article}

\usepackage[T1]{fontenc}
\usepackage{mathptmx}
\usepackage{microtype}
\usepackage{amsmath,amssymb}
\usepackage{mathtools}
\usepackage[a4paper,margin=2.5cm]{geometry}
\usepackage{authblk}

\usepackage{xcolor}
\usepackage{caption}
\usepackage{array,tabularx,ragged2e}
\usepackage[most]{tcolorbox}
\usepackage[scaled=0.97]{helvet}

\newtcolorbox[auto counter]{articlebox}[2][]{%
  enhanced,
  title={BOX~\thetcbcounter\hspace{3mm}#2},
  #1
}

\definecolor{BoxInk}{HTML}{293640}
\definecolor{BoxBlue}{HTML}{457990}
\definecolor{BoxMuted}{HTML}{647582}
\definecolor{BoxLine}{HTML}{D4E0E6}
\definecolor{BoxPale}{HTML}{F0F6F9}
\definecolor{BoxBody}{HTML}{FDFEFE}

\usepackage{hyperref}
\hypersetup{colorlinks=true,linkcolor=BoxBlue,citecolor=BoxBlue,urlcolor=BoxBlue}

\newcommand{\BoxSection}[1]{%
  \par\vspace{2pt}%
  {\color{BoxBlue}\sffamily\bfseries\fontsize{8.7}{11}\selectfont #1\par}%
  \vspace{3pt}{\color{BoxLine}\hrule height 0.35pt}\vspace{7pt}%
}
\newcommand{\TermRow}[2]{%
  \noindent\begin{tabularx}{\linewidth}{@{}>{\RaggedRight\arraybackslash\bfseries}p{0.245\linewidth}@{\hspace{5mm}}>{\RaggedRight\arraybackslash}X@{}}%
    #1 & #2
  \end{tabularx}\par\vspace{7pt}%
}

\title{Modeling Gene Expression Noise: From Transcriptional Dynamics to Cell Growth and Division}

\author[1]{Augustinas Sukys
    \thanks{Equal contribution.}}
\author[1,2]{Lucy Ham$^*$}
\author[1,2]{Kaan \"Ocal}
\author[3]{James Holehouse}
\author[4,5]{Ramon Grima%
  \thanks{Correspondence: \texttt{ramon.grima@stonybrook.edu}}}

\affil[1]{School of BioSciences, University of Melbourne, Parkville, Victoria 3010, Australia}
\affil[2]{School of Maths \& Stats, University of Melbourne, Parkville, Victoria 3010, Australia}
\affil[3]{Department of Biology, Washington University in St. Louis, St. Louis, Missouri, USA}
\affil[4]{Laufer Center for Physical and Quantitative Biology, Stony Brook University, Stony Brook, New York, USA}
\affil[5]{Department of Applied Mathematics \& Statistics, Stony Brook University, Stony Brook, New York, USA}

\date{}

\begin{document}

\renewcommand\Affilfont{\small}
\maketitle
\vspace{-2.5em}

\begin{abstract}
Gene expression is inherently noisy, leading to substantial cell-to-cell variation in mRNA and protein levels.
The telegraph model and the simpler bursty model provide good fits to single-cell count distributions and have been used extensively to understand transcriptional kinetics and estimate the kinetic parameters. However, these models all describe a static cell. Here, we review models of stochastic gene expression in growing and dividing cells which aim to understand how a cell’s own life history, from its birth, through copying its genome, to its division, leaves a distinct fluctuation signature on molecule numbers. We discuss the biochemical and biophysical processes captured by these models, their analytical tractability, the concept of cyclo-stationary behavior, when this is well approximated by the steady-state of static models and how the distributions of molecular numbers measured under these conditions sensitively depend on the experimental setup. This is followed by a discussion of how these models are fitted to single-cell measurements, how biological variability can be separated from measurement noise, and which computational advances make this feasible for thousands of genes. We finish by highlighting the open problems in the field.
\end{abstract}

\section{Introduction}

Gene expression is fundamentally noisy, and the resulting variability in molecular levels between cells has important consequences for cellular function
\cite{BAX2026240}. Modern single-cell technologies such as single-cell RNA sequencing (scRNA-seq) \cite{ziegenhain2017comparative} and single-molecule fluorescence in situ hybridization (smFISH) \cite{young2020technical}, together with fluorescent protein reporters \cite{urchueguia2021genome} and, increasingly, mass-spectrometry-based single-cell proteomics \cite{sabatier2025global,welter2026global}, are able to quantify molecule counts in individual cells. Combined with mathematical modeling, these measurements allow us to identify the sources of cell-to-cell variability, explain how it can be controlled or exploited, and clarify the role of noise in biological function.

\begin{figure}[htbp]
    \centering
    \includegraphics[width=0.95\textwidth]{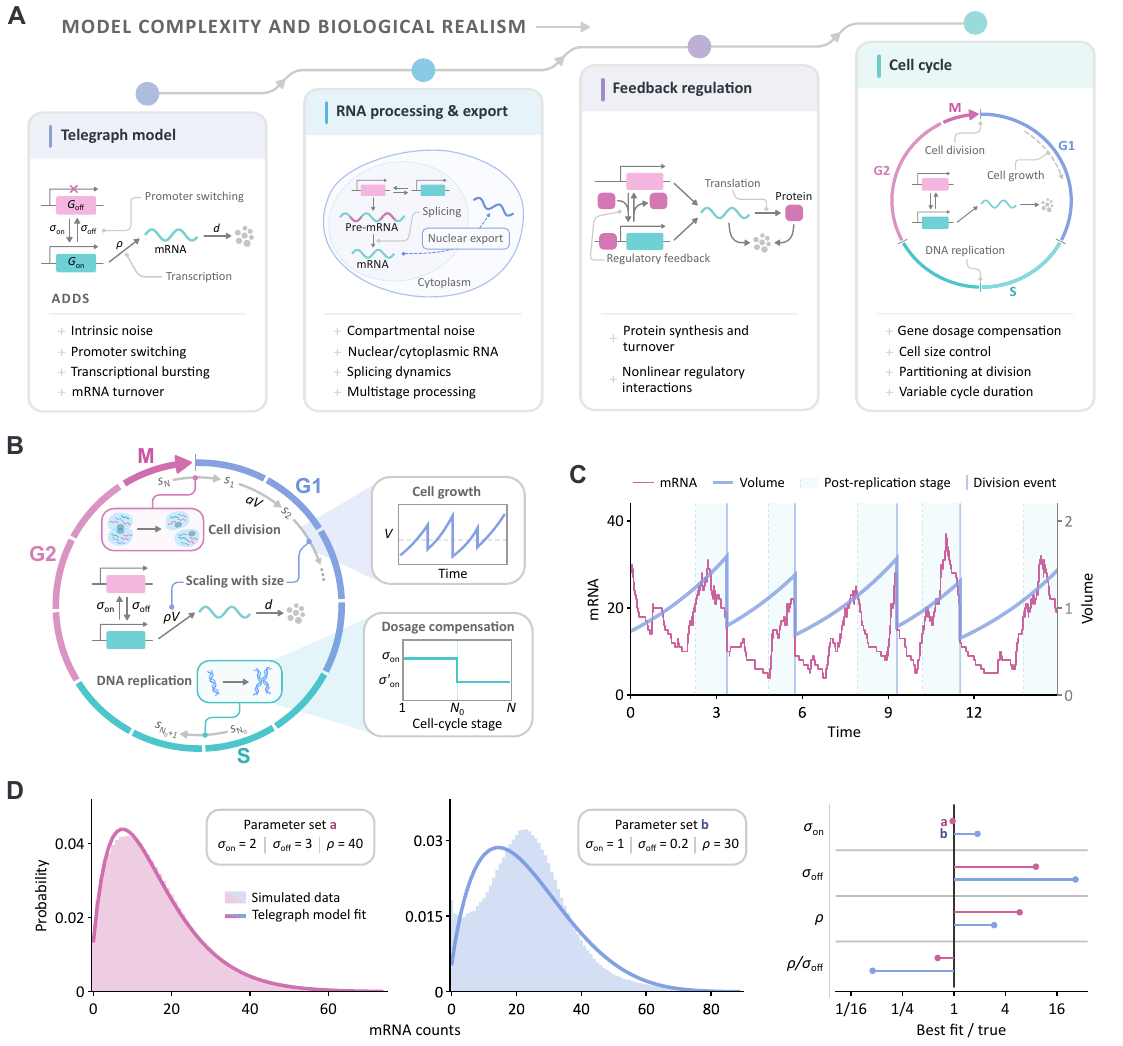}
    \caption{\textbf{Telegraph-model parameters are not accurately determined by data from growing and dividing cells.} \textbf{(A)} A hierarchy of model complexity. Extensions of the telegraph model incorporate RNA processing and nuclear export, feedback and cell-cycle dynamics. The schematics highlight the additional biological processes and sources of variability captured at each level. \textbf{(B)} A model of transcription that incorporates a number of cell-cycle effects neglected in the telegraph model, which combines elements of \cite{jia2022concentration,jia2023coupling}. For each gene copy, $G_{\mathrm{off}}\xrightarrow{\sigma_{\mathrm{on}}}G_{\mathrm{on}}$, $G_{\mathrm{on}}\xrightarrow{\sigma_{\mathrm{off}}}G_{\mathrm{off}}$, $G_{\mathrm{on}}\xrightarrow{\rho V}G_{\mathrm{on}}+M$, and $M\xrightarrow{d}\varnothing$; the cell-cycle stage advances as $s_k\xrightarrow{aV}s_{k+1}$ ($k=1,\dots,N$; $s_{N+1}\equiv s_1$), and the volume grows as $\dot V=gV$. Note that stage transitions scaling linearly with volume under exponential growth implement adder cell-size control (see Box~\ref{box:1}). On entry into $s_{N_0}$  the gene is replicated and $\sigma_{\mathrm{on}}\to\sigma_{\mathrm{on}}/2$ (perfect dosage compensation); on $s_N\to s_1$ the cell divides: $V\to V/2$, $M\to\mathrm{Bin}(M,1/2)$, and one gene copy is retained. \textbf{(C)} Stochastic simulation of a single lineage of the model in (B). Model parameters: $N=10$, $N_0=6$, $\sigma_{\mathrm{on}}=2$, $\sigma_{\mathrm{off}}=3$, $\rho=40$, $a=3$, $g=\ln 2/3$ (mean cycle duration $3/d$), rates in units of $d$. \textbf{(D)} Fitting the telegraph model using the method of maximum likelihood provides a good fit to the lineage mRNA distribution in some parameter regimes (left), but not in others (middle). In either case, the best-fit telegraph model parameter estimates differ greatly from the true values (right). The volume average is $\langle V \rangle \approx 1$, so the time-averaged transcription rates are approximately $\rho$. Model parameters are the same as in (C) for the left panel, and set to $\sigma_{\rm{on}}=1$, $\sigma_{\rm{off}}=0.2$, $\rho=30$ in the middle panel.}
    \label{fig:Figure1}
\end{figure}

The standard stochastic model of transcription is the telegraph, or two-state, model \cite{peccoud1995markovian}. It comprises four effective first-order reactions: promoter switching between inactive and active states, mRNA synthesis in the active state, and mRNA degradation (Figure~\ref{fig:Figure1}A; left). Its extension to proteins is the three-stage model, in which mRNA is translated and proteins are degraded \cite{shahrezaei2008analytical}. The telegraph model yields the mRNA-number distribution exactly, while the three-stage model can be solved approximately for the protein-number distribution when mRNA is much shorter-lived than protein. In suitable limits, both reduce to the bursty model. In this model, mRNA or protein is produced in geometrically distributed bursts, and the stationary distribution is negative binomial \cite{shahrezaei2008analytical}.

Both the telegraph model and the bursty model predict steady-state distributions of mRNA numbers that often fit snapshot data. Due to this good agreement, these models are often used to quantify transcriptome-wide bursting kinetics and characterise regulatory differences between genes \cite{ramskold2024single,tang_modelling_2023,grima2024quantifying,chen2026using}. When fitted to single-cell data, these models attribute variability between cells to fluctuations arising from the stochastic timing of biochemical events, commonly known as intrinsic noise. Because these models are analytically tractable, they also give direct quantitative insight.
For example, exact expressions can be derived for the steady-state entropy production rate of a leaky version of the telegraph model, quantifying how strongly transcription breaks detailed balance \cite{holehouse2026quantifying}.

The simplicity of these models comes at the cost of neglecting key biological detail. Transcription may involve multiple active or inactive promoter states, particularly in eukaryotes \cite{tunnacliffe2020transcriptional}, while effective first-order reactions cannot explicitly represent bimolecular interactions such as promoter interactions with transcriptional regulators and polymerase, or post-transcriptional regulation due to mRNA-microRNA interactions. Presently, these models also cannot simultaneously predict protein, nuclear mRNA counts and cytoplasmic mRNA counts per cell, even though their joint distribution is a natural quantity for describing gene expression. Furthermore, they assume the cell is static, i.e. they exclude a cell-cycle description involving birth, DNA replication, growth and division. These additional layers of biological detail can markedly affect molecule-count distributions. As a result, a good statistical fit of the steady-state solution of these models to single-cell data does not guarantee biophysically realistic predictions or meaningful kinetic parameters \cite{sukys2025cell,miles2025mechanistic}. Without correcting for these manifold effects, insights drawn from mathematical analysis of these models can be misleading. Figure~\ref{fig:Figure1} illustrates the problem. We simulated a lineage over many generations using a model with cell growth, DNA replication, dosage compensation, and division (Figure~\ref{fig:Figure1}B,C), and fit the resulting age-averaged mRNA distribution to the telegraph model (Figure~\ref{fig:Figure1}A; left). The fitting is good in some parameter regimes but not others; however, in both cases the estimates deviate substantially from the true values (Figure~\ref{fig:Figure1}D).

Recent experimental advances provide rich single-cell data that can be used to build more biologically realistic models (Figure~\ref{fig:Figure1}A; middle and right panels), making it possible to accurately infer kinetic parameters and to discriminate between competing mechanistic hypotheses. Live-cell imaging tracks the movement of polymerases, transcription factors and nascent RNA, thus providing an in-depth view of transcriptional dynamics \cite{eck2026single}; metabolic RNA labeling enables newly synthesized transcripts to be distinguished from pre-existing RNA, facilitating the separate estimation of RNA synthesis and degradation rates \cite{erhard2022time}; and multiomic assays jointly measure RNA with protein abundance \cite{xu2026simultaneous}.
Exploiting these richer measurements requires models that capture the underlying biology while remaining tractable for inference and mathematical analysis to facilitate insight. These demands motivate this review.

While the telegraph model has been extended to include multiple promoter states, explicit regulatory interactions and non-Markovian kinetics, our emphasis is on cell-cycle dynamics, which play a large role in shaping mRNA and protein distributions in cells. We focus on two developments: mechanistic models of gene expression in growing and dividing cells (Section~\ref{sec:growth_division}) and statistical and machine-learning methods that connect these stochastic models to single-cell data (Section~\ref{sec:inference}). Key terms are defined in Box~1.

\begin{figure}[p]
\begin{articlebox}[label={box:1}, width=\textwidth, sharp corners, boxrule=0.45pt, colframe=BoxLine, colback=BoxBody, colbacktitle=BoxPale, coltext=BoxInk, titlerule=0pt, left=4mm, right=4mm, top=1mm, bottom=2mm, toptitle=2mm, bottomtitle=2mm, fontupper=\fontsize{8.5}{10.4}\selectfont, coltitle=BoxBlue, fonttitle=\sffamily\bfseries\fontsize{8.7}{11}\selectfont]{\hspace{1mm} GENE EXPRESSION IN GROWING AND DIVIDING CELLS}
\setlength{\parindent}{0pt}
\renewcommand{\TermRow}[2]{\noindent\begin{tabularx}{\linewidth}{@{}>{\RaggedRight\arraybackslash\bfseries\hyphenpenalty=10000\exhyphenpenalty=10000}p{0.22\linewidth}@{\hspace{3mm}}>{\RaggedRight\arraybackslash}X@{}}#1 & #2\end{tabularx}\par\vspace{3pt}}
\vspace{2pt}
\BoxSection{CELL SIZE CONTROL}
\TermRow{Sizer / Timer / Adder}{A division event is triggered at a threshold size (sizer), once a fixed volume has been added to the birth size (adder), or at a fixed time after birth (timer).}
\vspace{2pt}
{\centering\includegraphics[width=0.9\linewidth]{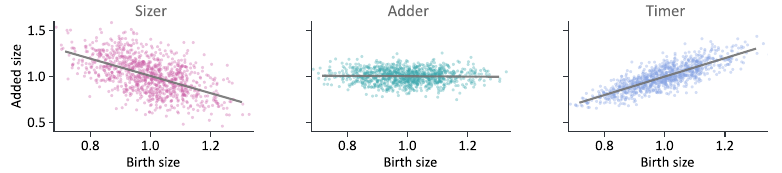}\par}
{\fontsize{8}{9.5}\selectfont Added size versus birth size (simulated): the slope identifies the cell-size control strategy \cite{vargas2018cell}.\par}
\vspace{4pt}
\BoxSection{CELL-CYCLE EFFECTS}
\TermRow{Size scaling}{Transcription per gene copy increases with cell volume which helps to maintain mRNA concentration homeostasis \cite{padovan2015single,sun2020size}.}
\TermRow{Dosage compensation}{DNA replication doubles gene copies; transcription per copy then drops, partially in mammalian cells \cite{padovan2015single,sukys2025cell}.}
\TermRow{Partitioning}{At division, mRNA is randomly partitioned between daughter cells \cite{baptista2020stochastic}.}
\vspace{2pt}
{\centering\includegraphics[width=0.9\linewidth]{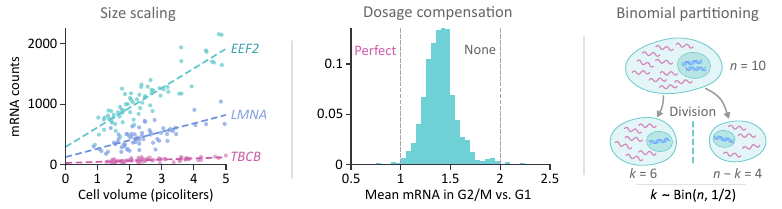}\par}
{\fontsize{7}{9.5}\selectfont Left: mRNA counts versus volume for \textit{EEF2}, \textit{LMNA}, and \textit{TBCB} genes in human fibroblasts (smFISH experiment), adapted from \cite{padovan2015single} with permission from Elsevier. Middle: ratio of mean mRNA counts in G2/M and G1 for 1760 bursty genes in mouse embryonic stem cells, adapted from \cite{sukys2025cell} under a Creative Commons license (CC BY 4.0).\par}
\vspace{4pt}
\BoxSection{STATIONARITY, CELL AGE AND SAMPLING}
\TermRow{Cyclo-stationary distribution}{Distribution of molecule numbers at a fixed cell age, after many generations, once it no longer depends on the generation.}
\TermRow{Age-averaged distribution}{The cyclo-stationary distribution of molecule numbers averaged over cell age, with the uniform age distribution for a lineage or the exponentially decaying age distribution of a growing population, in which young cells are over-represented.}
\vspace{2pt}
{\centering\includegraphics[width=0.9\linewidth]{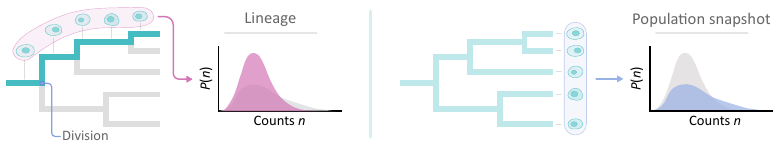}\par}
\end{articlebox}
\end{figure}

\section{Gene expression in growing and dividing cells}\label{sec:growth_division}

\begin{figure}[htbp]
    \centering
    \includegraphics[width=1\textwidth]{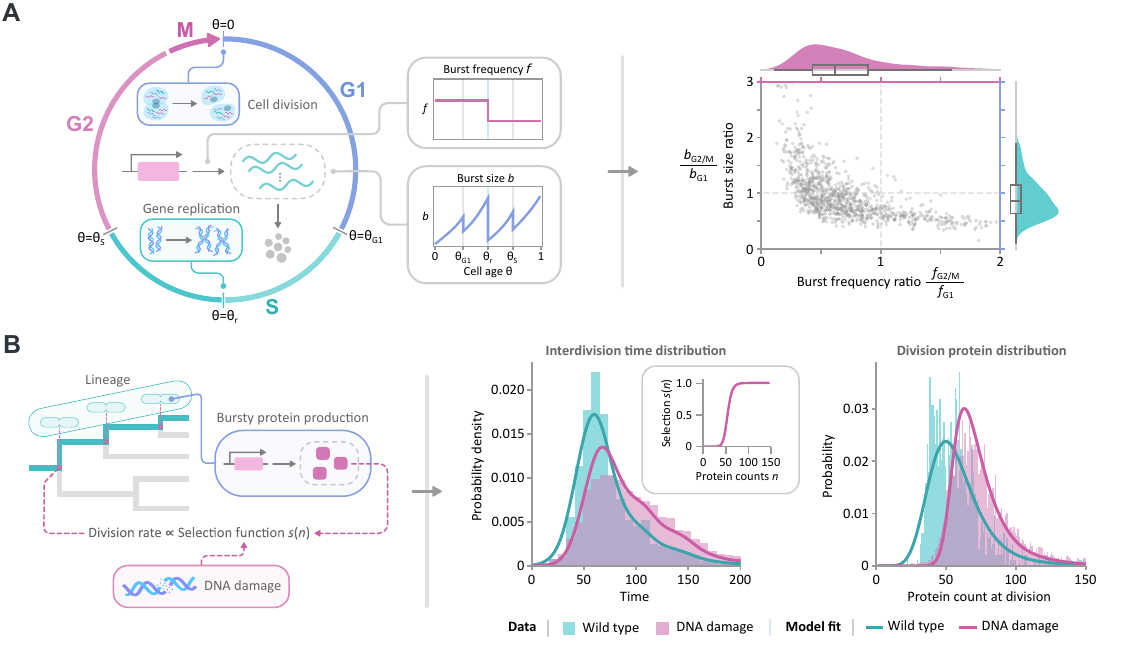}
    \caption{\textbf{Examples of modeling growing and dividing cells in the literature.} \textbf{(A)} Cell-cycle dependence of transcriptional bursting. Left: an age-dependent model accounts for gene replication, division and changes in burst frequency $f$ and burst size $b$ across normalised cell age $\theta$. Note that here the burst size is an exponential function of $\theta$ (with piecewise-constant parameters), which effectively captures the scaling of transcription with cell size. Right: fitting the model to cell-cycle resolved scRNA-seq data reveals a decrease in burst frequency with a comparatively stable burst size on the median level across the transcriptome, alongside substantial heterogeneity between genes. \textbf{(B)} Feedback between expression and division explains changes in the SOS response to DNA damage in \textit{E. coli}. Left: Bursty protein production is coupled to an age-dependent division rate governed by protein abundance $n$ through the selection function $s(n)$, with binomial partitioning at division. Right: Histograms show mother-machine lineage measurements of interdivision times and reporter abundance at division; solid curves show model fits. The inset shows the inferred selection function under DNA damage. Panels \textbf{A} and \textbf{B} are adapted from Refs. \cite{sukys2025cell} and \cite{piho_feedback_2024} under a Creative Commons license (CC BY 4.0).}
    \label{fig:Figure2}
\end{figure}

The telegraph model, like many of its extensions, including multiple promoter states \cite{jia2023analytical}, parameter variability across cells \cite{ham2020extrinsic} and feedback regulation \cite{kursawe2025efficient}, implicitly assumes a static cell that does not grow or divide (Figure~\ref{fig:Figure1}A). Many cells, like bacteria, yeast, cancer cells and cells mediating tissue renewal, however, proliferate. Cell growth and division add several sources of biological noise: random partitioning of molecules at division \cite{baptista2020stochastic}; the doubling of gene copy number upon DNA replication, partially offset by dosage compensation \cite{padovan2015single}; dependence of transcription parameters on cell size and cell-cycle stage  \cite{padovan2015single,sun2020size,fu2022quantifying}; variability in interdivision times and growth rates~\cite{soltani2016intercellular}, which can be affected by gene expression~\cite{piho_feedback_2024,stevanovic_dynamical_2024,levien2025slow} and in turn contribute to expression noise independently of fluctuations in protein production~\cite{kscheschinski2026single}. Cell size-dependent transcription is thought to be an important feature of eukaryotic transcription, as it helps to maintain stable mRNA concentrations during growth. In prokaryotes, by contrast, mRNA counts of some genes appear to change mainly at gene replication rather than with cell size \cite{kalita2025hfq}; nonetheless, gene expression noise can strongly affect microbial growth \cite{stevanovic_dynamical_2024,levien2025slow}. Illustrations of models with a number of the mentioned cell-cycle features are shown in Figure~\ref{fig:Figure1}B, Figure~\ref{fig:Figure2}A and B.

Since cell growth and gene expression are intrinsically coupled, fluctuations in one can affect the other, leading to long-term correlations across generations. All cells regulate their size, and the nature of cell size fluctuations sensitively depends on how this is done. Cell size control is commonly classified into three phenomenological strategies~\cite{vargas2018cell}: the \emph{sizer}, in which division occurs once a threshold size is reached; a \emph{timer}, in which it occurs a fixed time after birth; and an \emph{adder}, in which it occurs once a fixed volume is added since birth (Box~\ref{box:1}).
The three strategies can be distinguished, assuming exponential growth, by the slope of added size over the cycle against birth size. A sizer divides at a fixed size, so each extra unit at birth means one fewer added (negative slope); an adder adds a fixed amount regardless (zero slope); and a timer lasting one doubling time adds as much as it was born with (positive slope). These strategies are often applied to describe cell division timing, but can also be used to explain more general cell-cycle events \cite{cadart2018size,jia2022concentration}.

\textbf{Mechanistic models of gene expression.} Over the past decade, several studies have extended gene expression models to include cell growth and division~\cite{cao2020analytical,jia2023coupling,jia2022concentration,thomas_coordination_2021,perez2020effects,soltani2016intercellular,sun2019nonlinear,dessalles2020models,ali_cyclo-stationary_2025,wang2023exact,golding2024colloquium,baptista2020stochastic,biswas_cell_2026}. Results include moment equations for mRNA and protein numbers \cite{sun2019nonlinear,dessalles2020models,biswas_cell_2026}, as well as approximate or exact distributions of molecule numbers \cite{perez2020effects,jia2022concentration,thomas_coordination_2021,jia2023coupling,ali_cyclo-stationary_2025} and power spectra \cite{jia2022concentration}. Relatively few studies currently model the scaling of transcriptional activity with cell volume \cite{cao2020analytical,thomas_coordination_2021,jia2022concentration,jia2023coupling}. Including cell growth and division requires coupling gene expression to a model of the cell cycle, for example a multi-state Markov model of cell-cycle progression~\cite{jia2022concentration} or a noisy linear map relating size at division to size at birth~\cite{jia2023coupling}. These can be analyzed along a single lineage or, when population snapshots are of interest, incorporated into a branching process tracking all descendants \cite{thomas_coordination_2021}. Refs~\cite{jia2022concentration,jia2023coupling} also include DNA replication and dosage compensation, bringing the models closer to the biology of eukaryotic cells while still admitting analytical solutions. Cell-cycle dynamics have also been combined with modeling nascent RNA and nuclear export \cite{cao2020analytical}, protein production \cite{wang2023exact}, non-Markovian cell-cycle durations \cite{ali_cyclo-stationary_2025}, and feedback between gene expression and growth or division \cite{piho_feedback_2024,zhang2025stochastic,zabaikina2026gene}. Whilst the majority of studies have focused on the marginal distributions of mRNA or proteins in stationary conditions, recent work has also derived joint mRNA-protein number distributions or their correlations and investigated how they are affected by cell growth and division \cite{wang2023exact,biswas_cell_2026}. The growing complexity of these models has spurred the development in numerical methods, including a Finite State Projection for cell growth models \cite{jia2023coupling,piho_feedback_2024}, and agent-based stochastic simulation algorithms \cite{jia2022concentration, piho2026agentbasedmodeling,ocal_probing_2026}.

\textbf{Cyclo-stationary distributions.} In the presence of (almost) periodic events like cell division, the conventional concept of a steady state is not well defined. Nevertheless, in the long run, the behavior of these models is described by a cyclo-stationary state (see Box~\ref{box:1}), which describes e.g.~mRNA distributions conditioned on a fixed cell age (within the cell-cycle) or on the corresponding cell size. Averaging these cyclo-stationary distributions over the cell age distribution leads to age-averaged distributions which correspond to the experimental measurements.  Importantly, the age-averaged distributions differ when one considers a single lineage versus an entire population~\cite{jia2022concentration,ali_cyclo-stationary_2025,zhang2025stochastic}. This is because of different cell-age distributions -- growing populations are skewed towards younger cells, which typically have fewer molecules \cite{thomas2019intrinsic}. When fitting models to data, it is therefore crucially important to choose the right statistics (single lineage for mother machines, population for snapshots) to obtain correct parameter estimates.

\textbf{Relationship to conventional models.} Under certain conditions, the steady-state distribution of the conventional telegraph model with additional extrinsic noise (cell-to-cell variability in the parameters) exactly matches the age-averaged distribution of more complex cell-cycle models.
For example, \cite{thomas_coordination_2021,jia2023coupling} studied growing and dividing cells, where the transcription rate is proportional to the cell volume while disregarding DNA replication; they found that the
exact cyclo-stationary distribution at a given cell age $t$ and corresponding size $V$ is precisely the same as the steady-state solution of a telegraph model with a synthesis rate that is proportional to the cell size $V$ and an effective degradation rate equal to the sum of the (true) degradation and dilution rates. This implies equivalence between the age-averaged distribution of this model of growth and division, and the steady-state of a telegraph model with synthesis rates sampled from the distribution of cell volumes.
When DNA replication is modeled explicitly, Ref.~\cite{jia2023coupling} showed that the steady-state of the telegraph model with cell-dependent synthesis rate is a good approximation to the age-averaged distribution if two conditions hold: (i) the promoter is mostly inactive (bursty expression); (ii) dosage compensation upon DNA replication is strong. If either condition fails, the approximation can still hold when mRNA is fast compared to the cell cycle. In mammalian cells, some of these conditions are often violated: dosage compensation is, on average, of intermediate strength~\cite{sukys2025cell} (Box~\ref{box:1}), and the mean mRNA lifetime is not small compared with the mean cell-cycle duration~\cite{sharova2009database,schwanhausser2011global}.

\textbf{Key findings from modeling-based studies.}  (i) variability in division timing shifts levels of proteins and protein noise --- counterintuitively, as division timing becomes more stochastic, protein noise can decrease ~\cite{soltani2016intercellular}; (ii) neglecting variability in cell-cycle duration distorts the predicted mRNA variance most when the mRNA lifetime is comparable to the cell-cycle duration, and gene replication can generate bimodal mRNA distributions similar to those obtained from slow promoter switching~\cite{perez2020effects}; (iii) sizer-like control of the S phase, followed by timer-like control until division enables concentration homeostasis while reducing noise~\cite{jia2022concentration}; (iv) in multistable gene circuits, such as positive-feedback loops and toggle switches, isolated lineages switch reversibly between states, whereas in growing populations the faster-dividing state takes over, turning noise-driven switching into an irreversible cell-fate decision~\cite{piho_feedback_2024}; (v) when partitioning errors dominate, variable cell-cycle durations amplify noise in growing populations but not along single lineages, so lineage measurements can underestimate population noise~\cite{zhang2025stochastic}.

\section{Combining models and data}\label{sec:inference}

Most inference from single-cell data has thus far relied on conventional models which assume the existence of a steady-state: the bursty model \cite{grun2014validation}, the telegraph model \cite{larsson2019genomic,ramskold2024single,chen2026using} or its multiple state extensions (modeled explicitly \cite{trzaskoma20243d} or implicitly using non-Markovian models \cite{luo2023inferring}) or modifications of the bursty/telegraph models to account for cell-level parameter variability \cite{grima2024quantifying,tang_modelling_2023,gorin_monod_2025}. The accuracy of parameter estimation using these simpler models implicitly relies on how well their steady-state can approximate the age-averaged distribution of the more complex cell-cycle models, which is currently only partially understood. This is an important example of model misspecification, where the used model does not capture the underlying biological reality.

As illustrated in Figure~\ref{fig:Figure1}D, parameter estimation using the conventional telegraph model can be very misleading. Besides model mismatch, additional estimation error stems from technical noise which is particularly high for scRNA-seq because molecular capture probability per transcript is typically below $\approx 1/3$ \cite{10x_capture_2026} and varies considerably between cells and experimental protocols, making measurements of lowly expressed genes especially noisy. Estimation accuracy can be improved by accounting for technical noise by means of binomial thinning \cite{tang_modelling_2023,li2026efficiency}. However, even under ideal conditions, the errors in the estimated parameters can be considerable and vary strongly with the position in parameter space \cite{fu2022quantifying,gu_scalable_2025}; generally, the worst estimated parameter is the switching off rate \cite{fu2022quantifying}. When cell-to-cell variability becomes substantial, the telegraph model (and the bursty one) estimates will perform poorly, leading to overestimated burst sizes and underestimated burst frequencies \cite{grima2024quantifying}.

A more refined inference procedure involves identifying the major source of cell-to-cell variability (such as technical noise, cell size, nuclear processing machinery, gene cross-talk, cell-cycle phase or cell type) and accounting for it explicitly when fitting the telegraph model or one of its modifications \cite{fu2022quantifying,chari2023biophysically,grima2024quantifying,tang_modelling_2023,miles2025mechanistic,li2026prime,wang2025joint,zhou2026deciphering}. Nevertheless, all these methods assume a steady state, which as previously discussed can be a poor approximation to the solution of cell-cycle based models and limits our ability to discriminate between competing mechanisms~\cite{jiao2024can,nicoll2025transient}; models exploiting multimodal data partly alleviate the latter limitation \cite{wang2025joint}. The gold standard for overcoming these limitations is to fit mechanistic models of gene expression that explicitly incorporate cell growth and division to cell-cycle-resolved single-cell data~\cite{sukys2025cell,volteras_global_2024} (see, for example, Figure~\ref{fig:Figure2}A).

Parameters of the telegraph model and its extensions can be inferred by maximum-likelihood estimation \cite{gu_scalable_2025}; by minimising the Kullback--Leibler divergence between empirical distributions and model distributions obtained by Fourier inversion of the probability generating function (PGF) \cite{gorin_monod_2025}; by neural networks trained on stochastic simulations that emulate count distributions \cite{sukys2022approximating,huang2026deep}; and by fitting analytical PGFs directly to empirical PGFs, which matches the accuracy of FSP-based maximum likelihood at a fraction of the cost \cite{wang2025joint,li2026efficiency}. Parameters of mechanistic models of gene expression in growing and dividing cells have so far been inferred by maximum likelihood from cell-cycle-resolved snapshot data~\cite{sukys2025cell}, by maximum-caliber inference of path statistics~\cite{torres2023maxcal} and by approximate Bayesian computation matching summary statistics of time-resolved single-cell data~\cite{volteras_global_2024}.

\section{Conclusion and Outlook}\label{sec:conclusion}

The telegraph model and its bursty limit remain the most widely used descriptions of stochastic gene expression: they are simple, analytically tractable and empirically fit single-cell count distributions across many organisms and conditions. A good fit, however, does not mean that the inferred kinetic parameters are correct or even physically meaningful, as both models ignore the cellular life cycle. Proliferating cells grow, replicate their genome, modulate transcriptional kinetics with the cell cycle phase, partition their molecules at division and regulate their size; these processes strongly shape gene expression in ways not captured by the telegraph model. To understand cellular heterogeneity and noise in gene expression, we need models that couple gene expression to growth and division. These models allow us to determine when the static cell description of the  telegraph and bursty models provides a good approximation to the underlying biology, and to estimate how strongly the predicted distributions depend on measurement protocol.

Fitting these models to cell-cycle resolved single-cell data is a promising avenue to understand gene regulation at a genome-wide level, but poses several challenges. Snapshot distributions of a single species rarely discriminate between competing mechanisms of gene expression, and it is not yet clear which combinations of measurements (such as nascent and mature RNA, protein, multiple time points, cell age or size) are sufficient to determine the right mechanism.
A related difficulty lies in separating technical noise from common sources of extrinsic noise such as cell size --- achieving this at genome scale requires mechanistic models of gene expression and experimental designs in which most of the extrinsic variables are measured.
Beyond inference, the scope of current models needs to widen. Existing cell-cycle models assume simple promoter dynamics; extensions are needed to include multi-state promoters, regulatory interactions, nuclear processing and protein dynamics, which have mostly been studied without the cell cycle. These richer models will in turn add to the existing computational challenges in fitting mechanistic models to data. Exact likelihoods for cell-cycle models remain difficult to derive and computationally expensive to evaluate; it remains to be seen whether recent approaches developed for static cell models such as generating-function methods and neural emulators can be efficiently deployed for cell-cycle models.
Finally, spatial transcriptomics now enables the measurement of expression variability at the subcellular and tissue levels, but mechanistic models that combine stochastic gene expression with intracellular transport, cell-cell coupling and growth and division (especially important for malignant cells), and that can be fitted to such data, remain largely undeveloped.

Addressing these challenges would enable mechanistic models to do more than capture variability: they could predict gene expression responses to perturbations and new environments, link changes in noise to their consequences for cell physiology and fate, and predict the large-scale behavior of cell populations and tissues. Ultimately, the test of these models will be not only how well they fit existing data, but how well they anticipate new data.

\section*{Declaration of generative AI and AI-assisted technologies in the manuscript preparation process}

All first drafts of the writing were done without generative AI. Generative AI was used to assist the review and editing of human-written text, specifically for grammatical and typographical considerations.

\section*{Declaration of competing interest}
The authors have nothing to declare.

\section*{Acknowledgements}
L.H. acknowledges funding through the ARC Centre of Excellence for the Mathematical Analysis of Cellular Systems. K.Ö and A.S acknowledge funding through
an Australian Research Council Laureate Fellowship
(FL220100005). J.H. would like to acknowledge the support of the National Science Foundation Grant Award Number EF-2133863. R. Grima acknowledges support by the Laufer Center for Physical and Quantitative Biology. The authors thank Muhan Ma for helpful comments on an earlier draft.

\newcommand{\bibdoi}[1]{}

\makeatletter
\let\@refstar\@empty
\newcommand{\refstar}[1]{\gdef\@refstar{\textbf{#1}}}
\renewcommand{\@biblabel}[1]{%
  \settowidth{\@tempdima}{#1.}%
  \makebox[\labelwidth][r]{#1.\ifx\@refstar\@empty\else\llap{\raisebox{-2.2ex}[0pt][0pt]{\makebox[\@tempdima][c]{\@refstar}}}\fi}%
  \global\let\@refstar\@empty}
\newcommand{\refnote}[1]{%
  \\{\fontsize{\dimexpr\f@size pt-1.5pt\relax}{\dimexpr\f@baselineskip*4/5\relax}%
  \selectfont #1}%
}
\makeatother
\bibliographystyle{cosb-vancouver}
\bibliography{biorxiv}

\end{document}